\documentclass[preprint,prd,showpacs]{revtex4}
\usepackage{amssymb}
\usepackage{amsmath}

\begin{document}

\title{Energy spectrum of simply constant chromoelectric flux tubes}
\author{Shahin Mamedov ${}^{1,2}$}
\email{sh_mamedov@yahoo.com}

\author{Douglas Singleton ${}^{3,4}$}
\email{dougs@csufresno.edu}

\author{Shemsettin Turkoz ${}^{1}$}
\email{turkoz@science.ankara.edu.tr}

\affiliation{${\ }^{(1)}$ Department of Physics, Ankara University, Tandogan 06100, Ankara, Turkey \\
${\ }^{(2)}$Institute for Physical Problems, Baku State University, Z.Khalilov str. 23, Baku, AZ-1148, Azerbaijan \\
${\ }^{(3)}$ Physics Department, California State University Fresno, Fresno, California 93740-8031\\
${\ }^{(4)}$ Institute of Gravitation and Cosmology, Peoples' Friendship University of Russia, Moscow 117198, Russia}

\date{\today}

\begin{abstract}
In this article we obtain the energy spectrum of colored, spinor particles in
chromoelectric flux tubes. The chromoelectric field of the flux tubes considered
here comes from {\em simply} constant gauge potentials rather than from
covariantly constant gauge potential, as is usually the case. The energy
spectrum of the simply constant flux tubes is different than that of the covariantly constant
flux tubes. The spectrum is discrete due to the walls of the tube and with a
plus/minus constant shift depending on the magnitude of the constant chromoelectric background. This
goes against the classical intuition where one would expect a charged particle
in a uniform ``electric" field to accelerate with ever increasing velocity/energy
i.e. there would be no constant energy eigenvalue.

\end{abstract}


\maketitle

\section{Introduction}

Flux tubes -- both ``electric" and ``magnetic" -- play or are thought to play a role
in many physical systems: (i) In type II superconductors Abrikosov magnetic flux tubes
are an experimentally observed phenomenon \cite{abrikosov}; (ii) in Yang-Mills-Higgs theory Nielsen-Olesen flux
tubes are interesting theoretical solutions to the classical field equations \cite{nielsen}; (iii) in the dual
superconducting model of QCD, chromoelectric flux tubes are thought to lead to
quark confinement \cite{dual-super-cond}; (iv) in the glasma state which is thought to occur in
the early stages of high energy collisions of ultra heavy ions \cite{mclerran}.

In this article we find the energy spectrum of color charged, spinor particles inside a flux tube with
a constant chromoelectric field. The constant chromoelectric field we consider comes from
a constant gauge potential rather than usual case of a covariantly constant gauge potential
\cite{fuji}. There have been a few other works which have studied constant color
electric or color magnetic fields coming from simply constant vector potentials \cite{brown} \cite{reuter}.
The new features of the present work are: (i) we consider tubes of such
constant color fields, instead of having the constant color fields permeate all
space; (ii) we study the energy spectrum of spinor particles inside these tubes.
The energy spectrum is not the same as that produced by a covariantly
constant gauge potential, as is to be expected since the covariantly constant vector potential is
gauge inequivalent to the simply constant vector potential. The energy
spectrum of the simply constant potential has the unusual feature that the energy levels are those
of a particle inside a cylindrical tube but shifted up or down by some constant value
which depends on the strength of the chromoelectric field. Since the energy spectrum
is different for the two types of potentials which yield constant chromoelectric fields there might be some
situations where this could yields experimentally distinguishable results. One such possible
case might be the glasma state of matter which occurs in high energy heavy ion collision
such as at RHIC or ALICE. One could ask whether the flux tubes of the glasma state come from
covariantly constant potentials or simply constant potentials. In the conclusion we will
discuss the possible relevance of our results to the glasma state.

Previous work studying the instability of such simply constant field configurations
can be found in \cite{weis,agaev1,agaev2}. Related studies of the motion of
color charged particles in chromomagnetic and chromoelectric fields produced by
simply constant gauge potentials are given in the references
\cite{agaev1,weis2, mame1, mame2, mame3, fried, turk}.
In the present work we focus on the system of a spinor, colored particle in
a classical, chromoelectric flux tube produced by a simply constant gauge potential.
This problem is solved within the framework of ordinary quantum mechanics
i.e. we solve the energy eigenvalue problem of the
Dirac equation for a colored spinor in the classical background
of a simply constant vector potential. The confinement property of QCD is
taken into account by taking the wave function of the spinor particle
to vanish at the cylindrical walls of the flux tube.

\section{Quarks in the chromoelectric flux tube}

Quarks belong to the fundamental representation of the color group $SU_{c}(3)$. The Dirac equation for spinor
quarks in an external color field is obtained from the free Dirac equation by minimal coupling --
one shifts the momentum  as $P_{\mu }=p_{\mu }+gA_{\mu }=p_{\mu }+gA_{\mu }^{a}\lambda ^{a}/2$. This gives the Dirac equation for a quark
with mass $M$:
\begin{equation}
\label{1}\left( \gamma ^\mu P_\mu -M\right) \Psi =0,
\end{equation}
where $\gamma ^\mu$ are Dirac matrices, $A_{\mu}^a$ is the external gauge field,
$\lambda^a$ are the Gell-Mann matrices, $g$ is the color
interaction coupling constant, and the color index $a$ runs $a=1, ..., 8$.
The spinor  $\Psi $ has two Majorana components $\phi $ and $\chi $,
\begin{equation}
\Psi=\left(
\begin{array}{c}
\phi \\
\chi
\end{array}
\right).
\end{equation}
Each of these Majorana components are split into two spin components
\begin{equation}
\psi =\left(
\begin{array}{c}
\psi _+ \\
\psi _-
\end{array}
\right) ~,
\end{equation}
where $\psi $ stands for both spinors $\phi $ and $\chi$.
The components $\psi_{\pm} $ transform under the fundamental representation of the color group
$SU_c(3)$ and have three color spin components corresponding to the eigenvalues of color spin
operator $T^3=\lambda^3/2$:
\begin{equation}
\label{2} \psi _{\pm}=\left(
\begin{array}{c}
\psi _{\pm}\left(\lambda^3=1\right) \\
\psi _{\pm}\left(\lambda^3=-1\right) \\
\psi _{\pm}\left(\lambda^3=0\right)
\end{array}
\right)=\left(
\begin{array}{c}
\psi _{\pm}^{(1)} \\
\psi _{\pm}^{(2)} \\
\psi _{\pm}^{(3)}
\end{array}
\right) .
\end{equation}
Now we study of motion of quarks in a constant color electric
field coming from a {\em simply}, constant non-Abelian vector potential
as introduced in \cite{brown}. (Note in an Abelian theory it is not
possible to have constant electric or magnetic fields coming from a
constant vector potential). The field strength tensor for a
color electric/magnetic field coming from constant gauge potentials, $A_{\mu}^a$
is given by $F_{\mu \nu}^{c}=gf^{abc}A_{\mu }^{a}A_{\nu }^{b}$. The $f^{abc}$ are the
structure constants of $SU_{c}(3)$. The $i^{th}$
component of the color electric field corresponds to one of the indices
$\mu , \nu =0$ while the other index $\nu , \mu =i$. Without loss of generality we
will take the chromoelectric field to point in the $i=1$ or $x$-direction.

A constant chromoelectric field can be given by the following choice of
constant components of a non-Abelian vector potential $A_{\mu}^a$:

\begin{equation}
A_{\mu }^{1}=\left( \sqrt{\tau _{1}},0,0,0\right) ,\ A_{\mu }^{2}=\left( 0,%
\sqrt{\tau },0,0\right) ,\ \; \; \; {\rm all \; \; others} \; \;\; A_{\mu }^{a}=0,  \label{3}
\end{equation}
where $\tau $, $\tau_1 $ are constants. The field strength tensor $F^a_{\mu \nu}$ has one
non-zero component:

\begin{equation}
F_{01}^{3}=g\sqrt{\tau _{1}\tau }={\cal{E}}_{x}^{3}.
\label{4}
\end{equation}
As mentioned above the chromoelectric field is along the $x$-axis in ordinary space.
Also without loss of generality we have taken the chromoelectric field to point along
the $3$-axis in color space. We note again that
the constant color fields produced by covariantly constant as studied in \cite{fuji} are
gauge {\it inequivalent} to the simply constant potentials studied in \eqref{3}. Thus one
should be able to experimentally distinguish between these two options for obtaining
a constant color field flux tube.

In this background, chromoelectric field the squared Dirac equation has the form:
\begin{equation}
\label{5}
\left[  P^{\mu}P_{\mu} +
\frac{i~g}{4}\sigma^{\mu\nu}F^{c}_{\mu\nu}\lambda^{c}-M^{2} \right]
\Psi =0 ~,
\end{equation}
where $\sigma^{\mu\nu}=\frac{i}{2} [\gamma ^\mu, \gamma ^\nu]$. The
corresponding equations for the $\phi $ and $\chi$ spinors are:
\begin{equation}
\label{6}\left[  p^{2} -M^{2}
+\frac{1}{4}\left({\mathcal{G}}^{2}-{\mathcal{G}}_{1}^{2}\right)I_2+\mathcal{G}p_{1}\lambda
^{2} -{\mathcal{G}}_{1}E\lambda ^{1} \mp \frac{i ~
g}{2}\sigma^{x}{\cal{E}}^{3}_{x}\lambda^{3} \right] \phi,\chi =0.
\end{equation}
where here ${\mathcal{G}}= g\tau ^{1/2}$, ${\mathcal{G}}_{1}=$
$g\tau _{1}^{1/2}$ and
\begin{eqnarray}
I_{2}=\left(
\begin{array}{ccc}
1 & 0 & 0 \\
0 & 1 & 0 \\
0 & 0 & 0
\end{array}
\right)
\end{eqnarray}
is a color matrix. From \eqref{6}, only the $\psi _{\pm}^{(1,2)}$
states interact via the last three terms of the color interaction i.e.
\begin{equation}
\label{7}
\mathcal{G}p_{1}\lambda ^{2}
-{\mathcal{G}}_{1}E\lambda ^{1} \mp \frac{i ~
g}{2}\sigma^{x}{\cal{E}}^{3}_{x}\lambda^{3}\equiv \lambda^{a}I^{a}_{\mp}  ~,
\end{equation}
since $\left(\lambda ^a\right)_{i3} =0$ for $\psi_{\pm} ^{(3)}$ when $a=1,2,3$.
Here we have written the interaction term using the color vector $I^{a}_{\mp}$, 
which has components $\left(-{\mathcal{G}}_{1}E,\ \ \mathcal{G}p_{1}, \\  \mp 
\frac{1}{2}i~g\sigma^{x}{\cal{E}}^{3}_{x},\\ 0, 0, 0, 0, 0 \right)$. The interaction term is 
a conserved operator, since it commutes with equation \eqref{6} for the $\phi, \chi$ spinors. Squaring this operator
$\left( \lambda^{a}I^{a}_{\pm}\right)^2$ gives the following color diagonal form:
\begin{equation}
\label{8}
\left( \lambda^{a}I^{a}_{\pm}\right)^2=\left[
{\mathcal{G}}_{1}^{2}E^{2}+{\mathcal{G}}^{2}
p_{1}^{2}-\frac{1}{4}\left({\cal{E}}^{3}_{x}\right)^{2}\right]I_{2}.
\end{equation}
By separating the coupled equations in \eqref{6} one finds that $\psi
_{\pm}^{(1,2)}$ obey the same fourth order equation:
\begin{equation}
\label{9}
\left[ \left( E^{2}-{\bf{p}}^{2}
-M^{2}-\frac{1}{4}\left({\mathcal{G}}^{2}-{\mathcal{G}}_{1}^{2}\right)\right)^{2}-\left(
{\mathcal{G}}_{1}^{2}E^{2}+{\mathcal{G}}^{2}
p_{1}^{2}-\frac{1}{4}\left({\cal{E}}^{3}_{x}\right)^{2}\right)
\right]\psi _{\pm}^{(1,2)}=0 ~.
\end{equation}
Solving this last equation leads to two
relations between the energy/momentum of the spinor particle
and the parameters of the chromoelectric field:
\begin{equation}
\label{10}
 E^{2}-{\bf{p}}^{2}
-M^{2}-\frac{1}{4}\left({\mathcal{G}}^{2}-{\mathcal{G}}_{1}^{2}\right)=\pm\sqrt{{\mathcal{G}}_{1}^{2}E^{2}+{\mathcal{G}}^{2}
p_{1}^{2}-\frac{1}{4}\left({\cal{E}}^{3}_{x}\right)^{2}}.
\end{equation}
These relations are defined by the two eigenvalues of the $\lambda^{a}I^{a}_{\pm}$ operator
(i.e. $\pm\sqrt{{\mathcal{G}}_{1}^{2}E^{2}+{\mathcal{G}}^{2}
p_{1}^{2}-\frac{1}{4}\left({\cal{E}}^{3}_{x}\right)^{2}}$ ).
These two relations lead to the two branches of the continuous
spectrum \cite{weis2,agaev1}:
\begin{equation}
\label{11}
E^{2}_{1,2}={\bf{p}}^{2}
+M^{2}+\frac{1}{4}\left({\mathcal{G}}^{2}+{\mathcal{G}}_{1}^{2}\right)\pm
\left[{\mathcal{G}}_{1}^{2}\left({\bf{p}}^{2}+M^{2}\right)+{\mathcal{G}}^2
p_{1}^{2}\right]^{1/2}.
\end{equation}
The long range confinement property of QCD is dealt with as in
\cite{mame1} by taking the flux tubes to have a finite radius. This is done by
placing a cylindrical, hard wall at some radius, $r_0$. In principle we should also
place two hard walls at some positions along the $x$-axis. Doing this
does not really change the overall analysis so we not carry this out.
Note the operator ${\mathbf{p}}^{2}$
commutes with equation \eqref{9}. Thus we can solve the
eigenfunction/eigenvalue problem for ${\mathbf{p}}^{2}$ and then
combine these results with \eqref{9}. In cylindrical coordinates
with the axis of the cylinder along the $x$-axis the eigenvalue
equation for ${\mathbf{p}}^{2}$ is
\begin{equation}
\label{12}
{\mathbf{p}}^{2}\psi
^{(i)}_\pm\left({\mathbf{r}}\right)=-\nabla ^2\psi
^{(i)}_\pm\left({\mathbf{r}}\right)=p^2\psi
^{(i)}_\pm\left({\mathbf{r}}\right)
\end{equation}
For \eqref{12} the separation ansatz is $\psi^{(i)}_\pm\left(
\mathbf{r}\right) = R \left(r\right) \cdot u\left( \varphi \right)
\cdot \psi\left( x\right) \xi _{\pm }^{(i)}$ where $\xi _{\pm }^{(i)}$ contains the spin and
color spin parts of wave function and can be chosen as the eigenvectors of the corresponding $\sigma^3$ and $\lambda^3$ operators:
$$\xi _{\pm }^{(i)} = \xi_{\pm}\zeta^i ~,$$
where
$$\xi_{+}=\frac1
{\sqrt{2}}\left(
\begin{array}{c}
1\\0
\end{array}
\right), \\ ~ ~
\xi_{-}=\frac1
{\sqrt{2}}\left(
\begin{array}{c}
0\\
1
\end{array}
\right),
$$
and
$$
\\ \zeta^1=\frac1
{\sqrt{3}}\left(\begin{array}{c}
1\\0\\0
\end{array}
\right), ~~ \zeta^2=\frac1
{\sqrt{3}}\left(\begin{array}{c}
0\\
1\\0
\end{array}
\right),\ ~~ \\ \zeta^3=\frac1
{\sqrt{3}}\left(\begin{array}{c}
0\\0\\1
\end{array}
\right).
$$
The separation ansatz divides \eqref{12} into three independent equations:
\begin{equation}
\label{13}
\left[\frac 1r\frac \partial {\partial r}\left( r
\frac{\partial }{\partial r}\right) +\left(
p^2-p_1^2-\frac{m^2}{r^2}\right) \right]R \left( r\right) =0
\end{equation}
\begin{equation}
\label{14}
-\frac{\partial ^2 }{\partial \varphi ^2}u\left( \varphi
\right)=m^2u\left( \varphi \right)
\end{equation}
 \begin{equation}
\label{14a}
 \left[\frac{\partial ^2 }{\partial x^2}+p_1
^2\right]\psi\left( x\right) =0 ~,
\end{equation}
where $r=\sqrt{y^2+z^2}$. The solution of \eqref{14} is $u\left(
\varphi \right) =\frac 1{\sqrt{2\pi } }e^{im\varphi }\ \left(
m=0,\pm 1,\pm 2,...\right) $. Here $m$ is the chromomagnetic quantum
number and is defined as the projection of the chromomagnetic moment
onto the field direction i.e. the axis of the cylinder or the $x$-axis.
The chromomagnetic moment is connected with the orbital motion of
colored particles \cite{mame1,mame3}. Next, the eigenvalue equation \eqref{14a} has the
general solution:
\begin{equation}
\label{15}
\psi\left( x\right)= B e^{i p_1 x} ~,
\end{equation}
where $p_1$ is the momentum of the particle along the $x$-axis.
Finally the solutions to \eqref{13}, under the requirement that the wave function
be finite at the origin, are Bessel functions of the first kind
\begin{equation}
\label{17}
R\left( r\right) =C J_m\left( p_{\bot }r\right),\
\end{equation}
where $p_{\bot }=\sqrt{p^2-p_1^2}$ is the transverse momentum. The
interaction of the chromomagnetic moment of the quarks with the
background field is characterized by the index $m$. However, the
energy spectrum \eqref{11} does not depend on this quantum number
and thus, up to this point, we have a degeneracy over the quantum
number $m$.

For the radial direction we still need to take into account the boundary condition that the wave function of the quark
should vanish as the cylinder wall i.e.
\begin{equation}
\label{18}
R\left( r=r_0\right) =C J_m\left( p_{\bot }r_0\right)=0.\
\end{equation}
This yields the condition $p_{\bot} r_0 = \alpha _m^{(N)}$ where
$\alpha _m^{(N)}$ is the $N^{th}$ zero of the $m^{th}$ Bessel function of the
first kind. This gives the following quantization condition on $p_{\bot}$:
\begin{equation}
\label{19} 
\left(p_{\bot }\right)_m^{(N)}=\frac{\alpha_m^{(N)}}{r_0}.\
\end{equation}
Here $N$ is the radial quantum number. Inserting the quantized
value transverse momenta \eqref{19} and the free momentum $p_1$ along the
$x$-axis  into the spectrum \eqref{11} we obtain the energy spectrum for the quark:
\begin{equation}
\label{spectrum}
 \left(E^{(N)}_m \right)^{2}_{1,2}=  p_1 ^{2}+\left( \frac{\alpha _m^{(N)}}{r_0}\right)^2 +M^{2}+\frac{1}{4}\left({\mathcal{G}}^{2}+{\mathcal{G}}_{1}^{2}\right)
\pm \sqrt{{\mathcal{G}}_{1}^{2}\left( p_1 ^{2}+\left( \frac{\alpha _m^{(N)}}{r_0}\right)^2+M^{2}\right)+{\mathcal{G}}^{2} p_1 ^{2}}.
\end{equation}
The energy spectrum \eqref{spectrum} now depends on $m$ through $\alpha _m^{(N)}$
and thus the interaction with the boundary eliminates the degeneracy of the energy spectrum
over the $m$ quantum number.
As noted earlier the states $\psi_{\pm}^{(3)}$ do not interact with the color background thus
the spectrum for these states is simply \cite{mame1} :
\begin{equation}
\label{20} 
\left( E_m^{(N)}\right) _3^2=p_{\bot}^2+p_1^2+M^2.
\end{equation}
However, these states do interact with the boundaries thus the solutions
for the $\psi_{\pm}^{(3)}$ states are like those for the
$\psi_{\pm}^{(1,2)}$ states but without the color background
term which shifts the energy states up/down. Imposing the flux tube boundary conditions for the
$\psi_{\pm}^{(3)}$ states we get the quantization condition
\eqref{19}. Inserting these into \eqref{20} gives the
spectrum for the $\psi_{\pm}^{(3)}$ states
\begin{equation}
\label{21}
 \left( E_m^{(N)}\right) _3^2=\left( \frac{\alpha_m^{(N)}}{r_0}\right) ^2+  p_1 ^2 + M^2.
\end{equation}
This spectrum for the $\psi_{\pm}^{(3)}$ states is completely independent of the
color fields/color interaction terms

One can simplify the energy spectra in \eqref{spectrum} with the following
two assumptions: (i) We take $\tau = \tau_1$ and thus ${\cal G} ={\cal G}_1 = g \sqrt{\tau}$. (ii)
For light quarks ($u,d,s$) the mass, $M$ can be neglected with respect
to the QCD field strength/energy scale set by $g \sqrt{\tau}$.
With these assumptions \eqref{spectrum} becomes
\begin{equation}
\label{spectrum2}
 \left(E^{(N)}_m \right)^{2}_{1,2}=  p_1 ^{2}+\left( \frac{\alpha _m^{(N)}}{r_0}\right)^2 +\frac{1}{2} g^2 \tau
\pm \sqrt{2 g^2 \tau p_1 ^{2}+\left( \frac{\alpha _m^{(N)}}{r_0}\right)^2} ~.
\end{equation}
The energy spectrum is split into three
distinct levels: (i) An upper level, $(E^{(N)}_m)_1$; (ii) a middle level
$(E^{(N)}_m)_3$; (iii) a lower level $(E^{(N)}_m)_2$.
The splitting into the three levels given by \eqref{spectrum2} and \eqref{21}
is due to the chromoelectric field. This has similarities to the splitting of the $n=2$ level of
hydrogen by an ordinary electric field in the Stark effect -- of the four $n=2$, electron energy levels
one is shifted up, one is shifted down and two are un-shifted. The middle level,
$(E^{(N)}_m)_3$, comes simply from the confining boundary conditions, while the upper and
lower levels, $(E^{(N)}_m)_{1,2}$, are split due to the constant chromoelectric field.
Again this type of behavior is not what one would expect for a constant ``electric" field which
should accelerate the ``charged" particle continuously and have now energy eigenvalue. 

\section{Spectrum of emission in the chromoelectric flux tubes}

The transition between the different energy levels given by \eqref{21} and \eqref{spectrum2} can occur either
via gluons or photons since the quarks carry both color and electric charge. However since the characteristic
time scale for the electromagnetic interaction is  many orders of magnitude longer ($\approx 10^{-16}$ seconds)
as compared to the strong interaction ($\approx 10^{-24}$ seconds) the transitions will occur essentially only via gluons
which will manifest themselves as jets.

The explicit expressions for the energy spectrum of quarks in a
chromoelecric flux tube given in \eqref{21} and \eqref{spectrum2}
may be studied by measuring the spectrum of gluons (jets) emitted
by this tube. The energies of the emitted gluons are given by
\begin{equation}
\label{gluon-spec}
\omega=E_{m', n'}^{\left( N' \right)}- E_{m, n} ^{\left(N\right)} ~,
\end{equation}
where $N', m', n'$ are the quantum numbers of the initial state and
$N, m, n$ are the quantum numbers of the final state. There are many
different permutations of initial and final quantum numbers, but there
are some restrictions coming from conservation of certain operators.
First, the color vector $\lambda^a I^a_{\pm}$ of \eqref{7} is a conserved quantity as discussed in
section II. The states corresponding to the three different energy
branches, $E_{1,2,3}$, have different values of $\lambda^a I^a_{\pm}$ and so transitions
between these three branches are forbidden by conservation of this operator.
Second there is the restriction that $\Delta m = m'-m = 0, \pm 1$
for transitions between the levels within each of the three branches, $E_1$, $E_2$, $E_3$. This is the usual selection rule
associated with the operator $L_z$ which is the conserved component of angular momentum
along the special axis of the problem. In this case the special direction is called $x$
and so the operator is actually $L_x$.

By studying the spectrum of emitted gluons (which will manifest as jets), 
\eqref{gluon-spec}, and taking into account the
selection rules associated with the operators $\lambda^a I^a_{\pm}$ and $L_x$, one should
be able to observe the three energy branches $E_1, E_2$ and $E_3$. In order for the emitted
gluons (jets) to be observable they must have energies  in the $10 \times$GeV range so that
one is in the pertubative regime of QCD. The energy scale between adjacent
levels is set by the radius of the flux tube, $r_0$. Taking $r_0$ to be of a size typical
of strongly interacting systems namely $1$ GeV$^{-1}$ then implies that the energy between
adjacent levels is of order $1$ GeV. In order to have the emitted gluons have energies of
$10 \times$ GeV or greater means we are looking for transitions where
the final $N$ and/or $n$ are of the order $1$ while the initial $N'$ and/or $n'$ should
be of order $10$ or larger. Since the quantum number $m$ is restricted by $\Delta m = m'-m = 0, \pm 1$
it can not give the needed splitting between final and initial energy levels.
Having $N'-N \ge 10$ would give a difference between the energy levels of
$10 \times$ GeV or larger and the emitted gluon would appear as a jet. As mentioned in section II there is a
gauge inequivalent way of obtaining constant color fields via covariantly constant
non-Abelian vector potentials \cite{fuji}. The energy spectrum of such constant color fields
coming from covariantly constant non-Abelian vector potentials would be different
from the energy spectrum obtained here and one could experimentally distinguish between
these two types of chromoelectric flux tubes.

\section{Discussion and conclusion}

The central result of this paper is the explicit energy spectrum
given in \eqref{spectrum} for a spinor, color charged particle
(i.e. a quark) inside a chromoelectric flux tube of radius $r_0$,
and its application to obtaining the spectrum of emitted gluons (jets) in
\eqref{spectrum2} \eqref{21} \eqref{gluon-spec}. This result may
have applications to the glasma state \cite{mclerran} which is
thought to occur during the initial stages of heavy ion collisions
such as those studied at RHIC or to be studied at ALICE. As the
heavy ions collide and pass through one another, chromoelectric
and chromomagnetic flux tubes are thought to form. The fields of
these color flux tubes of the glasma state are assumed to behave
like classical, color field flux tubes. Thus the results presented
here might be used to probe for the presence of these color field
flux tubes by looking for gluons/jets which exhibit the discrete
spectrum of \eqref{spectrum2} \eqref{21} \eqref{gluon-spec}. For
a flux tube with a non-confining boundary the continuous spectrum
\eqref{11} will be useful. In this case for comparison with
experiments the energy of the moving quark jet should be measured
directly.

The spectrum given by \eqref{spectrum} depends on several quantum number, $n, m, N$, and
as such yields a complex spectrum for emitted gluons/jets \eqref{gluon-spec}. The spectrum is restricted by
two selections rules: (i) No transitions occur between the three different energy
branches, $E_{1,2,3}$ due to the conservation of the color projection operator $\lambda^a I^a_{\pm}$.
(ii) Due to conservation of the angular momentum projection operator, $L_x$, one has the
standard selection rule $\Delta m = 0, \pm 1$ between the energy levels of
each branch, $E_{1,2,3}$. There is also a practical requirement that the observable
transitions are those which occur between widely spaced levels. For example, the initial
$N'$ and/or $n'$ should be of order $10$ or greater and the final $N$ and/or $n$ should be of order
$1$. This requirement comes about since for the gluons to appear as jets they should
be in an energy range where QCD is perturbative i.e. they should have energies
of the order or greater than $10 \times$ GeV rather than 1 GeV.  This spectrum for a quark in the chromoelectric flux tube is
similar to molecular spectrum with several quantum numbers (electronic, rotational, and vibrational
quantum numbers) rather than the simple hydrogen atom spectrum with its single, radial quantum
number. At present our results already allow us to say that the spectrum
of gluons from the glasma state might be complex and the energy of the emitted gluons
would be in the $10 \times$ GeV range or larger. By experimentally measuring
the gluon/jet spectrum and thus the energy spectrum one could determine physical characteristics
of the glasma such as the chromoelectric field strength ${\cal{E}}^{3}_{x} =
g \sqrt{\tau \tau _1}$.

In this article we have only investigated chormoelectric flux
tubes. In practice the chromomagnetic flux tubes arise at the same
time with chromoelectric ones \cite{mclerran}. One can calculate
the energy spectrum of the chromomagnetic flux tubes in a manner
exactly similar to the procedure used in this article for
chromoelectric flux tubes and one will arrive at a spectrum
similar to \eqref{spectrum}. However in the case of a color
electric charged quark placed in a color magnetic flux tube one
finds that this system develops a field angular momentum, ${\bf
L}_{QCD}$ coming from the color electric and color magnetic
fields. This field angular momentum is given by \cite{ji}
\begin{equation}
\label{qcdam}
{\bf L} _{QCD} = \frac{1}{4 \pi} \int {\bf r} \times ({\bf E}^a
\times {\bf B}^a) d^3 x ~,
\end{equation}
where ${\bf E}^a$ and ${\bf B}^a$ are the color electric field of the quark
and color magnetic field of the flux tube. This field angular momentum coming from a color
electric charge in combination with a color magnetic dipole was
conjectured to play a role \cite{sing} in explaining the
``missing" spin of the proton \cite{emc}. The relevant color
fields in regard to calculating the QCD field angular momentum of
\eqref{qcdam} are a longitudinal cylindrical chromomagnetic field
and Coulomb chromoelectric field. The field angular momentum of
the electromagnetic version of such a system (longitudinal
magnetic field and Coulomb electric field) was calculated in
\cite{sing2} where it was found that ${\bf L}_{EM}$ was
proportional to the radius of the flux tube, the strength of the
magnetic flux and the strength of the electric charge. In the
approximation we have used in this paper, where the fields are
treated classically, the QCD field angular momentum will also have
the same kind of dependence (i.e. ${\bf L}_{QCD} \propto r_0, g$).
This field angular momentum will shift the total angular momentum
of the system of chromomagnetic flux tube plus quark. Naively the
angular momentum of this system should just come from the quark,
but taking \eqref{qcdam} into account means the system will have a
total angular momentum of ${\bf L}_{total} = {\bf L}_{quark} +
{\bf L}_{QCD}$ where ${\bf L}_{quark}$ is the spin plus orbital
angular momentum of the quark \cite{ji}. Additionally, the field
angular momentum of \eqref{qcdam} can have an indirect influence
on the energy spectrum of the chromomagnetic flux tube plus quark
through the introduction of an additional restriction on the
parameters that determine the energy spectrum. In this work the
color fields have been treated as classical fields and therefore
the field angular momentum as well is classical. However, it is
known that quantum mechanically angular momenta must be quantized
in integer steps of $\hbar /2$, thus we should have ${\bf L}
_{QCD} = \hbar ~ n' /2$ where $n'$ is an integer. This condition
gives a further quantization of the parameters, $r_0, g$, which
are used in calculating ${\bf L}_{QCD}$ (this is the same
procedure used in deriving the Dirac quantization condition for
monopoles \cite{saha}). Thus in addition to the shift of the
angular momentum of the chromomagnetic flux tube plus quark system
due to \eqref{qcdam}, the energy spectrum will have an additional
restriction due to the quantization condition ${\bf L} _{QCD} =
\hbar ~ n' /2$. The investigation of the details of the
chromomagnetic flux tube plus quark will be the topic of a future
paper.

\section*{Acknowledgments}

This work has been done under grant 2221 of TUBITAK organization (Turkey).
Sh.M. acknowledges members of High Energy Group of Ankara University for interest to
this work and hospitality. D.S. is supported by a 2008-2009 Fulbright
Scholars grant. D.S. would like to thank Vitaly Melnikov for
the invitation to research at the Center for Gravitation and Fundamental
Metrology and the Institute of Gravitation and Cosmology at PFUR.


\begin{thebibliography}{99}

\bibitem{abrikosov}
A.A. Abrikosov, Sov. Pys. JETP, {\bf 5}, 1174 (1957)

\bibitem{nielsen}
H.B. Nielsen and P. Olesen, Nucl. Phys. B {\bf 61}, 45 (1973)

\bibitem{dual-super-cond} Y. Nambu, Phys. Rev. D {\bf 10} 4262 (1974); S. Mandelstam,
Phys. Rept. {\bf 23}, 245 (1976); S. Mandelstam, In
Proc. 1979 Int. Sym. on Lepton and Photon Interactions at High Energy (ed. T.B.W. Kirk, and
H.D.I. Abarbanel). Fermilab, Batavia, Illinois (1979)


\bibitem{mclerran}
H. Fujii and K. Itakura, Nucl. Phys. A {\bf 809}, 88 (2008) (arXiv:0803.0410 [hep-ph]);
T. Lappy and L. McLerran, Nucl. Phys. A {\bf 772}, 200 (2006) (arXiv:0602189 [hep-ph]);
Larry McLerran, ``Strongly Interacting Matter at High Energy Density"
(arXiv:0812.1518 [hep-ph])

\bibitem{fuji} H. Fuji, K. Itakura and A. Iwazaki ``Instabilities in
non-expanding glasma" (arXiv:0903.2930 [hep-ph])

\bibitem{brown}  L. S. Brown and W. I. Weisberger, Nucl. Phys. B {\bf 157}, 285 (1979)

\bibitem{reuter} M. Reuter and C. Wetterich, Phys. Lett. B {\bf 334}, 412 (1994);
G.C. Nayak and P. van Nieuwenhuzien, Phys. Rev. D {\bf 71} (2005) 125001

\bibitem{weis}   W.I. Weisberger, Nucl. Phys. B {\bf 161}, 61 (1979)

\bibitem{agaev1} Sh.S. Agaev, A.S. Vshivtsev, V.Ch. Zhukovsky, P.G. Midodashvili,
Vestn. Mosk. Univ., Fiz. {\bf 26}, No. 3 12-16 (1985)

\bibitem{agaev2}  Sh.S. Agaev, A.S. Vshivtsev, V.Ch. Zhukovsky, Yad. Fiz. 36
(1982) 1023; Sh.S. Agaev, A.S. Vshivtsev, V.Ch. Zhukovsky, O.F. Semyonov,
Sov.Phys.J.28 (1985) 67; Sh.S. Agaev, A.S. Vshivtsev, V.Ch. Zhukovsky, O.F.
Semyonov, Sov.Phys.J. {\bf 28}, 29 (1985)

\bibitem{weis2}  R. Akhoury and W.I. Weisberger, Nucl. Phys. B {\bf 174}, 225 (1980)

\bibitem{mame1}  Sh. Mamedov, Eur. Phys. J.\ C {\bf 30}, 583 (2003) (arXiv:
hep-ph/0306179)

\bibitem{mame2}  Sh. Mamedov, Eur. Phys. J.\ C {\bf 33}, 537 (2004) (arXiv:
hep-ph/0310273)

\bibitem{mame3}  Sh. Mamedov, Eur. Phys. J.\ C {\bf 49}, 983  (2007)
(arXiv: hep-th/0608142)

\bibitem{fried} M.H. Friedman, Y. Srivastava and A. Widom, J.Phys. G {\bf 23}, 1061 (1997)

\bibitem{turk} S. Turkoz, Sh. Mamedov and A.K. Ciftci, Cent. Eur. J. Phys. B(3) 2010, 378


\bibitem{iwaz} A. Iwazaki, ``Pair Creation of massless fermions in electric flux
tube", arXiv:0905.2003 [hep-ph].

\bibitem{ji} X. Ji, Phys. Rev. Lett. {\bf 78}, 610 (1997); I. Balitsky
and X. Ji, Phys. Rev. Lett. {\bf 79}, 1225 (1997)

\bibitem{sing} D. Singleton, Phys. Lett. B {\bf 427}, 155 (1998)

\bibitem{emc} J. Ashman {\it et. al.}, Phys. Lett. {\bf B206},
364 (1988); Nucl. Phys. {\bf B328}, 1 (1989)

\bibitem{sing2} D. Singleton and J. Dryzek, Phys. Rev. B {\bf 62}, 13070 (2000);
J. Dryzek, {\it et. al}, Int. J. Mod. Phys. D {\bf 11}, 417 (2002)

\bibitem{saha} M.N. Saha, Ind. J. Phys. {\bf 10}, 145 (1936);
.ibid Phys. Rev. {\bf 75}, 1968 (1949); H.A. Wilson, Phys. Rev. {\bf 75},
309 (1949)


\end{thebibliography}
\end{document}